\documentclass{isprs} %

\usepackage{subfig}
\usepackage{breakcites}
\usepackage{geometry} %
\usepackage{epstopdf}
\usepackage[labelsep=period]{caption}  %
\usepackage[english]{babel}
\usepackage{url}
\usepackage{microtype}
\usepackage{natbib}
\usepackage{acronym}
\usepackage{booktabs}
\usepackage{amsmath}
\usepackage{balance}
\usepackage[svgnames]{xcolor}

\begin{document}

\newacro{lstm}[\textsc{lstm}]{Long Short Term Memory}
\newacro{gdd}[\textsc{gdd}]{Growing Degree Days}
\newacro{conab}[\textsc{conab}]{Companhia Nacional de Abastecimento}
\newacro{ndvi}[\textsc{ndvi}]{Normalized Difference Vegetation Index}
\newacro{csm}[\textsc{csm}]{Crop Simulation Models}
\newacro{svm}[\textsc{svm}]{Support Vector Machines}
\newacro{rf}[\textsc{rf}]{Random Forest}
\newacro{nn}[\textsc{nn}]{Neural Network}
\newacro{evi}[\textsc{evi}]{Enhanced Vegetation Index}
\newacro{sif}[\textsc{sif}]{Solar-Inducted Chlorophyll Fluorescence}
\newacro{mape}[\textsc{mape}]{Mean Average Percent Error}
\newacro{ibge}[\textsc{ibge}]{Brazilian Institute of Geography and Statistics}
\newacro{pam}[\textsc{pam}]{Produção Agrícola Municipal}
\newacro{sgd}[\textsc{sgd}]{Stochastic Gradient Descent}
\newacro{rmse}[\textsc{rmse}]{Root Mean Squared Error}

\title{ESTIMATING CROP YIELDS WITH REMOTE SENSING AND DEEP LEARNING{\color{white}\thanks{
    \expandafter\color{red} Author pre-print. Accepted for publication at LAGIRS 2020.
}}}

\author{
	Renato Luiz de Freitas Cunha, Bruno Silva
}

\address{
    IBM Research, São Paulo, Brazil\\ \texttt{\{renatoc,sbruno\}@br.ibm.com}
}

\abstract{
    Increasing the accuracy of crop yield estimates may allow improvements in
    the whole crop production chain, allowing farmers to better plan for
    harvest, and for insurers to better understand risks of production, to name
    a few advantages. To perform their predictions, most current machine
    learning models use NDVI data, which can be hard to use, due to the
    presence of clouds and their shadows in acquired images, and due to the
    absence of reliable crop masks for large areas, especially in developing
    countries.  In this paper, we present a deep learning model able to perform
    pre-season and in-season predictions for five different crops. Our model
    uses crop calendars, easy-to-obtain remote sensing data and weather
    forecast information to provide accurate yield estimates.
}

\keywords{Deep Learning, Remote Sensing, NDVI, Yield Estimation, Modeling}

\commission{}{}
\workinggroup{}
\icwg{}

\maketitle

\sloppy
\section{Introduction}\label{sec:introduction}

In 2050, the world's population is expected to reach 9.7
billion~\citep{un2019}, it represents approximately 2 billion more people in
the next 30 years. To feed all this population, an active increase in
agricultural productivity is key to fight a potential food gap
\citep{fao2009}. Technology plays a key role in this subject by providing new
techniques to increase farming productivity including yield forecasts that
can be used to plan strategic ways to perform agricultural management
activities.

Accurate crop yield forecasts are essential in decision-making for the food
industry, farmers, and governments~\citep{wang2018deep,
you2017deep}. It helps farmers in planning activities related to crop harvest,
storage, and distribution, while also improving the efficiency of government
resource allocation, mainly in developing countries. Additionally, a precise yield forecast improves the
decision-making process with regards to imports and exports of agricultural products.

Currently, most yield forecast techniques employ at least one remote sensing
data source as a feature for yield prediction. A very common approach is the
usage of \ac{ndvi} or other satellite data bands. These bands are then
combined with previous yields to build models of future yields. \ac{ndvi} is
obtained by the composition of near-infrared and red spectral channels. As
these vegetation indexes are obtained from direct observation over the crop,
they provide high-quality insights related to plant health almost in
real-time. However, There are two main drawbacks when using \ac{ndvi} to
predict yield: (i) the planting should be executed prior to \ac{ndvi}
acquisition, and (ii) for large regions it is hard to obtain a
reliable crop calendar definition and to determine where each crop was
planted (i.e., crop mask) especially in developing countries.

Another important approach for yield prediction corresponds to the
utilization of \ac{csm} such as \textsc{dssat}, \textsc{wofost},
\textsc{pcse}, and \textsc{apsim}~\citep{jin2018review}. These models usually
require the utilization of four data inputs: weather, genetics, soil, and
farm management. For a single farm, this kind of solution works well,
as genetic and management data are relatively simple to obtain. In large
regions, however, this solution can be expensive, with acquisition of local
data from many different farmers becoming a challenge.

In this paper, we propose a deep learning model for pre-season and in-season
crop yield estimation using data from easy-to-obtain data sources. As we do not
have to process \ac{ndvi} data, our solution is easily scalable and can present
predictions for large regions in a very fast way. We only use geographic
coordinates, weather and soil data, and crop calendars to
perform yield estimates. Because Brazil is a country with great agricultural
output, we consider five major Brazilian crops in this study: soybean, corn,
rice, sugarcane, and cotton.
The contributions of this work are the following:

\begin{itemize}
    \item A solution for yield forecast that demands fewer data when compared
    to existing methods that require large amounts of remote sensing data.
    These solutions are not feasible for large regions where an accurate,
    annotated crop map does not exist. Our solution retrieves the input
    parameters from lightweight input sources and requires only weather and
    soil properties for a given latitude and longitude.

    \item We provide a fast and scalable yield forecast model. The provided
    model does not need crop masks indicating what, when, and where crops
    were planted. Therefore, our model does not need to process large amounts
    of data to provide a yield prediction.

    \item Pre-season yield forecasts. Although our proposed solution can perform
    in-season yield forecasts, it can also predict yield before seeding.
    Therefore, farmers can be more prepared to take management decisions like
    rental of machinery, people allocation and price negotiation.

    \item Region specific crop calendar. The proposed model utilizes
    customized weather data according to region-specific crop calendar
    without changing the model feature set. In this way, we do not need
    multiple models for different regions, improving model prediction power
    and scalability.

\end{itemize}

The paper is organized as follows. Section \ref{sec:related} presents the
related work. Section \ref{sec:model} shows the proposed model. In Section
\ref{sec:experiments}, we discuss the experiments and results followed by the
conclusion in Section \ref{sec:conclusion}.
\section{Related work}\label{sec:related}

Yield prediction systems have been widely used over the last decades not only to
provide important insights to farmers about potential crop productivity, but
also to serve social needs such as (i) food security, (ii) policy assessment,
(iii) yield gap analysis, and (iv) resource usage and
efficiency~\citep{holzworth2015agricultural,louhichi2010generic}.  In this
section, we present current efforts in the field, highlighting studies that
focus on data-driven approaches. 

\cite{kogan2013winter} assessed the efficiency of winter wheat yield
predictions in Ukraine using three different methods. Initially, \ac{ndvi} data
from \textsc{modis} and \textsc{esa} GlobCover maps were employed to feed
linear regression models and provide predictions in a 250m spatial
resolution. They also employed an empirical model based on meteorological
observations using data from 180 weather stations collected over 13 years.
Finally, WOFOST crop growth simulation model ~\citep{boogaard1998wofost} was
adopted in the yield forecasting process. This method simulates the
biophysical processes that occur between plants and the environment, which
involves algorithms for representing phenology, canopy development, biomass
accumulation, water stress, and many other plant development processes. The
different yield estimation methods were evaluated in a 2--3 months period and
\textsc{wofost} provided the best results using the \ac{rmse} metric.

\cite{cai2019integrating} proposed a combined approach using satellite and
climate data to predict wheat production in Australia. The authors performed a
series of experiments comparing traditional methods such as regression and
machine learning approaches (\ac{svm}, \ac{rf}, and \ac{nn}). They used two
sources of satellite data: (i) \ac{evi}, and (ii) \ac{sif}. As climate data
inputs, the authors employed several variables, including precipitation,
temperature, and solar radiation. The study concludes that the combination of
climate and satellite data can achieve higher performance when compared to
satellite-only for the studied region and period. Although the results
suggest the combination of satellite and climate may lead to better results,
in practice these results can be only achieved if a reliable crop calendar
and annotated farm locations are available. 

A multi-task learning algorithm which exploits spatial and temporal features
for predicting yield in cotton fields is presented by
\cite{nguyen2019spatial}. Different from other approaches that consider a
homogeneous behavior within crop fields, the authors model spatial variations
in the current field for soil, climate, tillage, and irrigation conditions
and the neighbors' potential correlation in yield assessment. The authors
compared the proposed approach with conventional machine learning methods
such as linear regression, decision tree regression, \ac{rf}, \ac{svm},
 and \textsc{xgb}oost. For the evaluated fields, the proposed approach
outperforms the results of other conventional methods.

\cite{wang2018deep} employ transfer learning to predict soybean
yield in Argentina and Brazil from a limited dataset. They used a trained model
with Brazilian data and augmented the model with Argentine data. The results
seem to be promising as the authors were able to get better results in the
Argentine model via transfer learning. They also were able to improve the
existing Brazil model with a transfer-learning model trained on Argentine
and Brazilian data. To perform crop yield prediction, the authors leveraged
different \textsc{modis} imagery products including eight-day composites for
seven-band reflectance imagery, two-band daytime/nighttime temperature
imagery, and a land cover mask.

The work presented by \cite{oliveira2018scalable} employs a neural network to
predict yield for corn and soybean using weather and soil data sources. Our
proposed model extends that model by
(i) adding new crop types, (ii) making the model structure dependent on crop
type, and (iii) adding new features to the model, such as \ac{gdd}.

Our solution differs from previous work as it requires only soil and weather
data to predict yield for large regions. The proposed model can also perform
predictions before planting. We also employ different
crop calendars in the same model to consider the planting/harvest
characteristics of each evaluated region.
\section{Model}\label{sec:model}

\begin{figure*}[!ht]
    \begin{center}
        \includegraphics[width=1\linewidth]{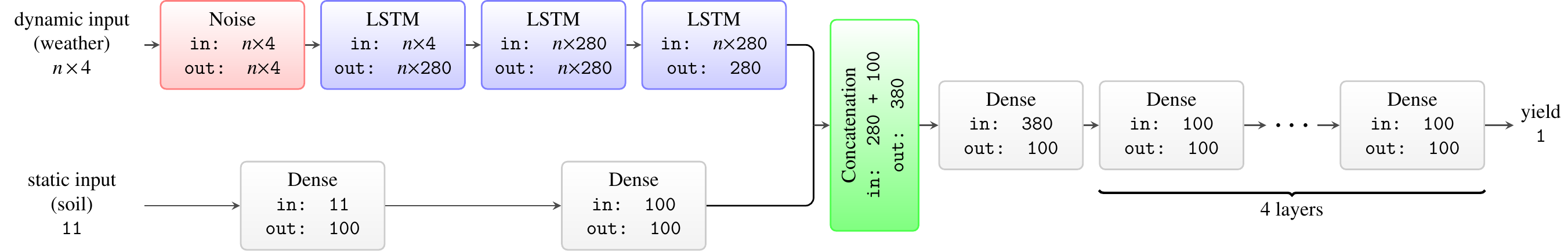}
    \end{center}
    \caption{%
        Proposed deep neural network architecture. The Red node corresponds
        to a noise generation layer. Blue nodes represent Long Short-Rerm
        Memory (LSTM) recurrent layers. Gray nodes represent dense,
        fully-connected layers. The green node represents a concatenation
        layer, which concatenates the intermediate representations from the
        dynamic and static paths. Numbers below node names represent shapes
        of input and output tensors. For example, $x \times y$ represents a
        $x$ by $y$ matrix, while single numbers represent line vectors. $n$
        corresponds to the crop cycle length (in months). This figure was
        extended from \cite{oliveira2018scalable}.
    }\label{fig:architecture}
\end{figure*}

In this section, we describe the model (Figure~\ref{fig:architecture}) used
to solve the problem of estimating yields of the studied crops. We also
describe the data sources used and the features added to the model. Our model
is based on the one proposed by~\citet{oliveira2018scalable}. The model has
two separate data paths (dynamic and static) that merge inside the model. The
rationale behind this design decision is that, in doing so, so-called dynamic
data, such as weather data, can be processed, and specialized, by different
nodes than static data, such as location and soil data. More specifically,
time-series data such as weather forecasts can be processed by
\ac{lstm}~\citep[][1997]{hochreiter1997long} nodes, which tend to work well
with time-series data, while soil data can be processed by fully-connected
nodes.

Our model expands on the original model~\citep{oliveira2018scalable} by (i)
adding accumulated \ac{gdd} as a dynamic feature alongside weather forecasts;
(ii) making the length of the dynamic features dependent on the type of crop
being used by using crop calendars as input; (iii) including an additive
zero-centered Gaussian noise node at the input of the dynamic data path.

Incorporating accumulated \ac{gdd} as a feature may help the model better
account for the influence of temperature on the crops studied. This decision
is justified because different crops have different \ac{gdd} values. For
example, corn requires around 1600-1770$^\circ$\textsc{c} \ac{gdd} for achieving full-season
maturity~\citep[][1987]{neild1987nch}, while soybeans require around
1300-1500$^\circ$\textsc{c} \ac{gdd} from planting to physiological
maturity~\citep{george1990effect}. In providing this feature, we expect the
model to have more opportunity to learn meaningful mappings from features to
predicted yield.

One potential flaw of the model proposed by~\citet{oliveira2018scalable} is
that they use the same time window length for both crops studied (namely,
soybeans and corn). In doing so, they risk adding more (or less) data to the
model, harming performance by making data selection harder for
the model. To mitigate this problem, we incorporated domain knowledge of crop
development into our model. This data comes from crop calendars, which
indicate typical planting and harvesting dates for crops in a given region.
In Brazil, this information is published by
\ac{conab}~\citep{mendes2019calendario}.

Training a model that uses weather data as a feature has an inherent
challenge: if the weather data comes from observations, the model won't be
exposed to uncertainty when used with weather forecasts. Additionally, even
if weather forecasts are input as points, the model will not have been
exposed to uncertainty in the weather forecasts because those would be input
as point estimates. Therefore, we've introduced a regularization layer that
adds zero-centered noise to the normalized dynamic inputs of the neural
network, which doubles as a random data augmentation method.
\section{Experiments}\label{sec:experiments}

\begin{figure*}[htbp]
    \centering
    \subfloat[Corn\label{fig:2a}]{\includegraphics[width=0.2\textwidth]{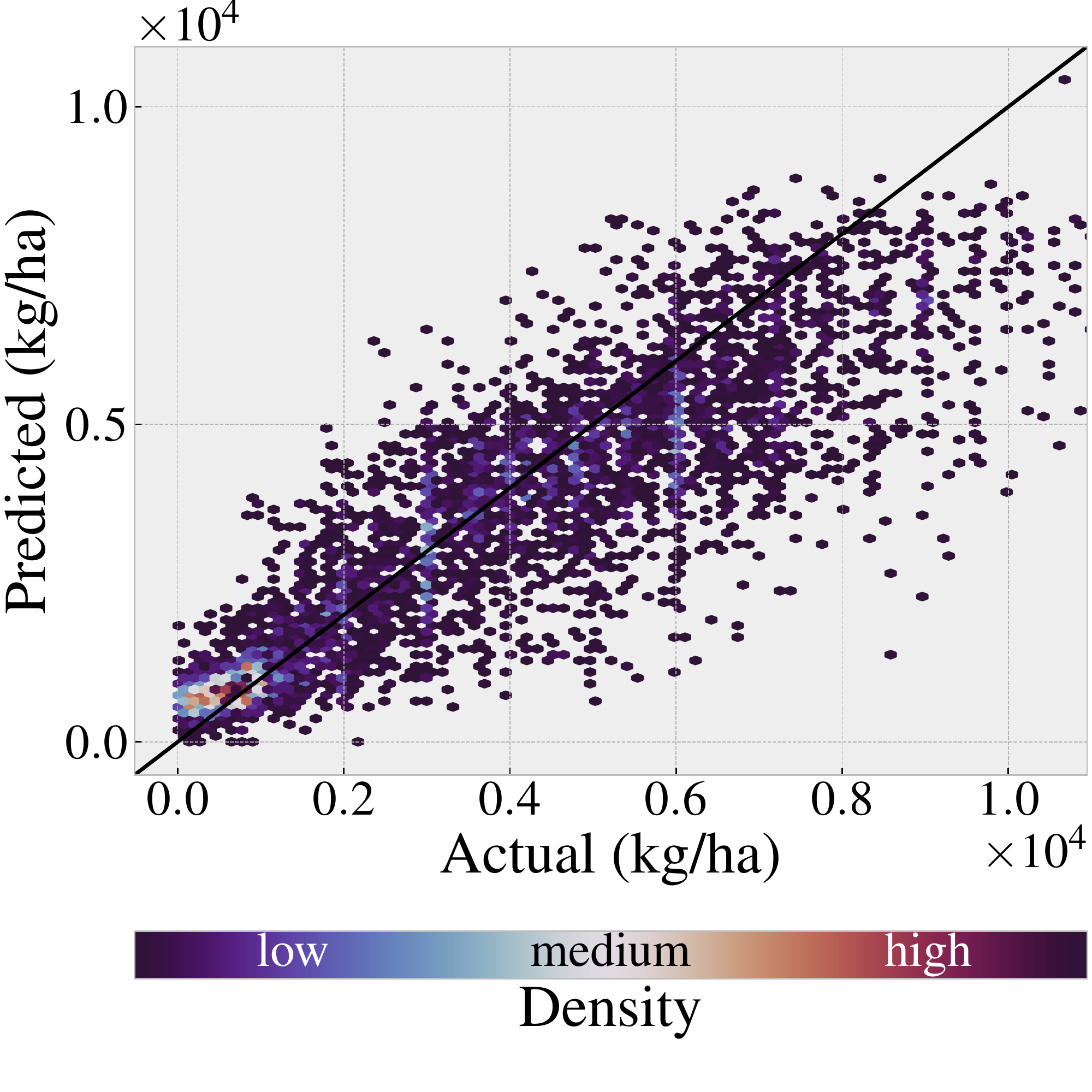}}\hfill
    \subfloat[Cotton\label{fig:2b}]{\includegraphics[width=0.2\textwidth]{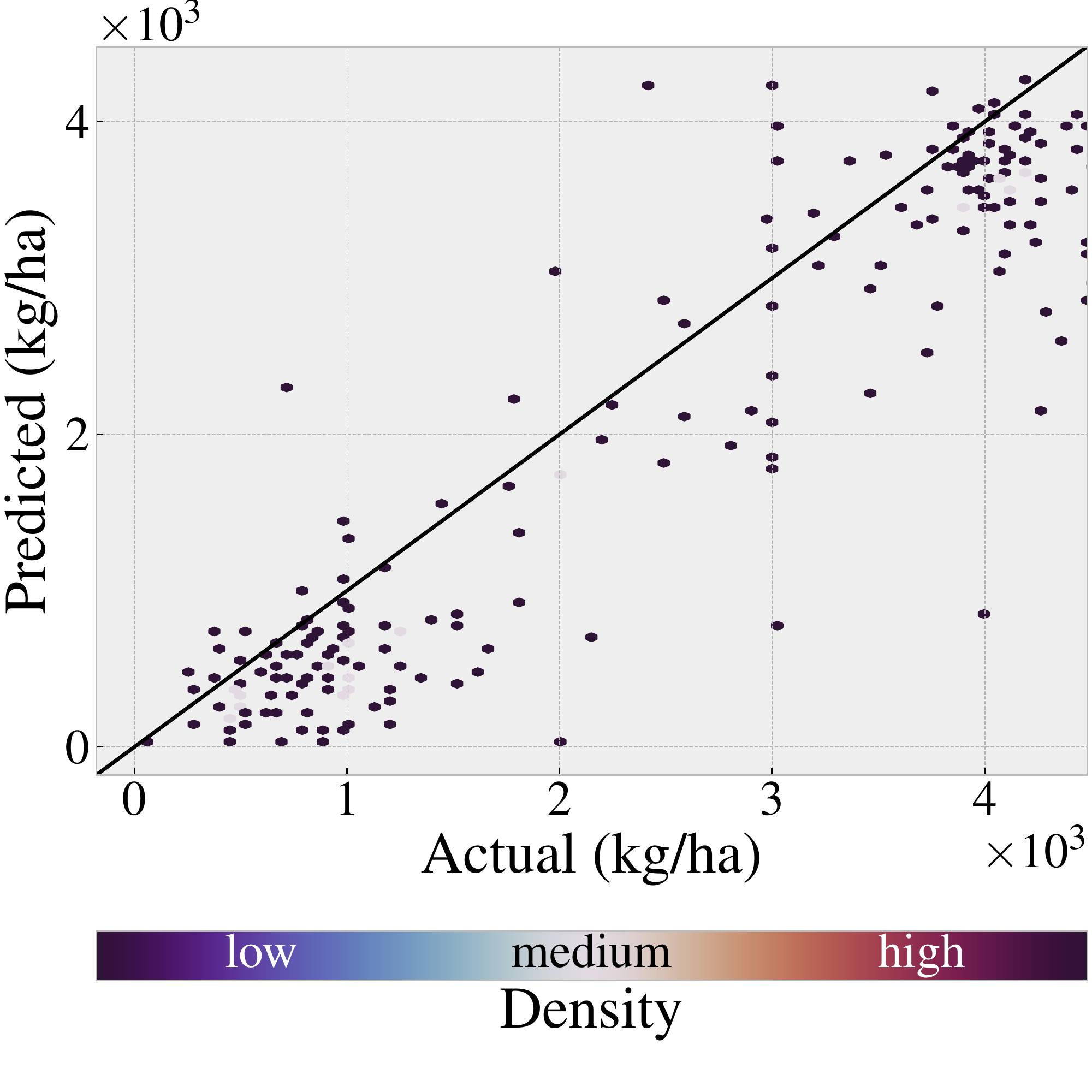}}\hfill
    \subfloat[Rice\label{fig:2c}]{\includegraphics[width=0.2\textwidth]{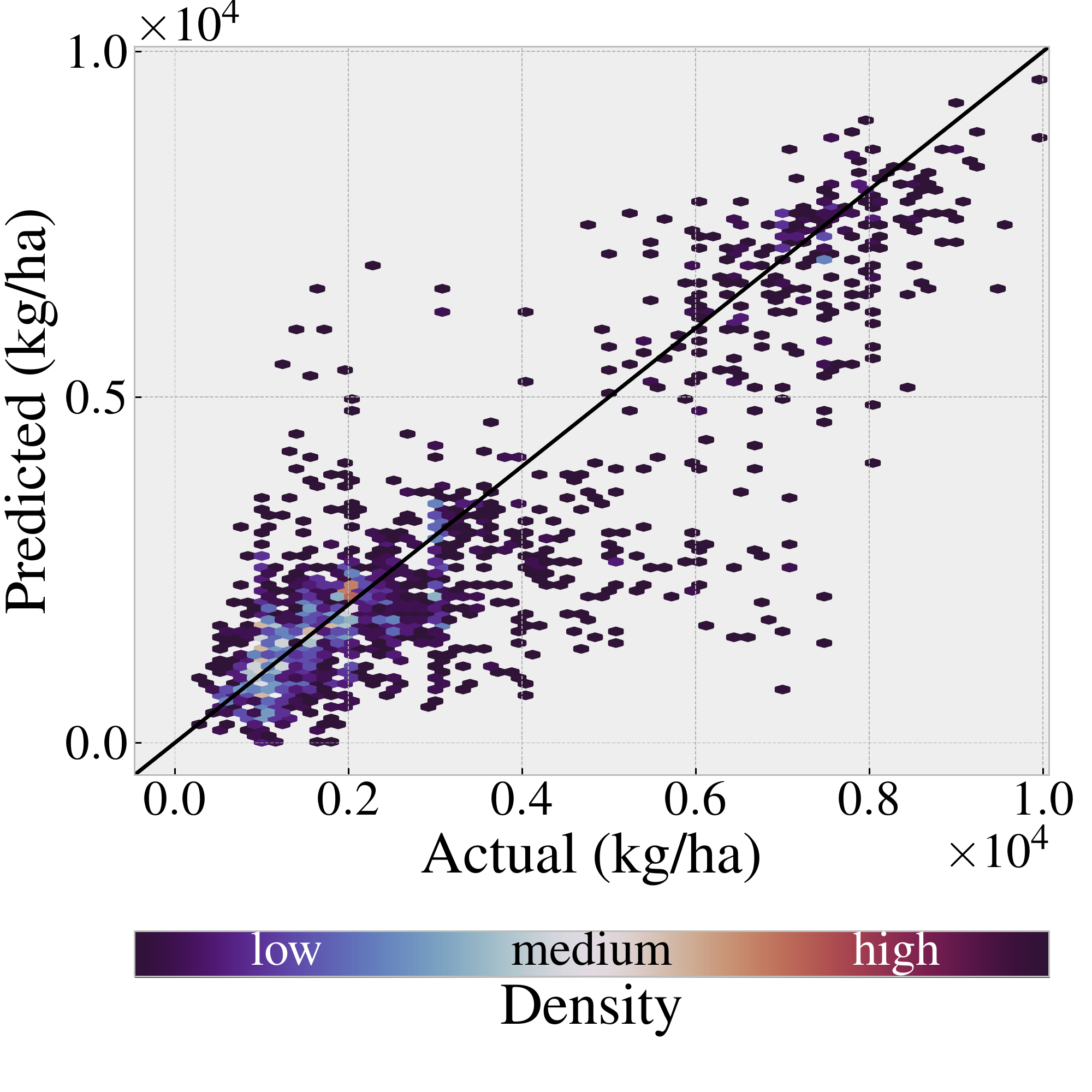}}\hfill
    \subfloat[Soybean\label{fig:2d}]{\includegraphics[width=0.2\textwidth]{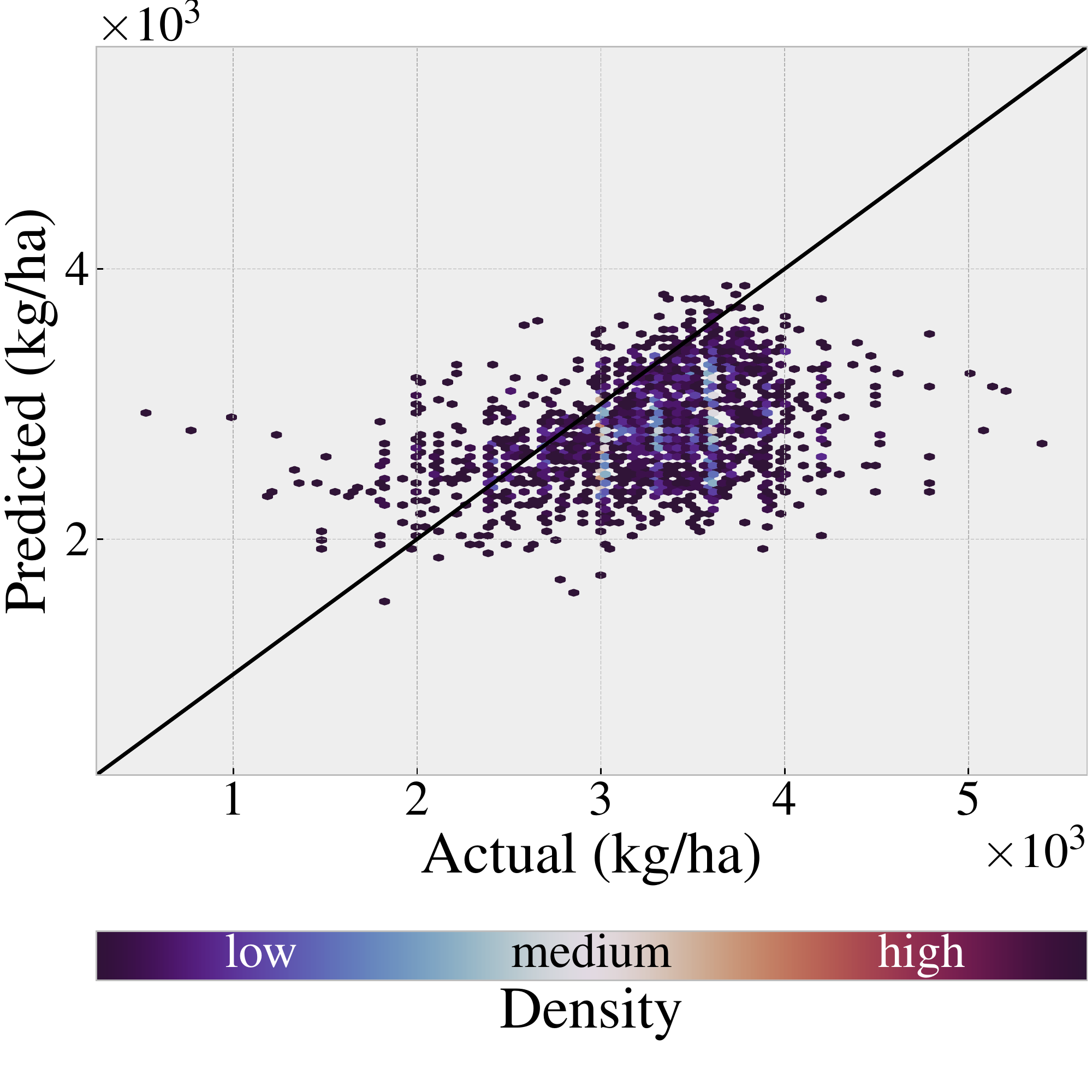}}\hfill
    \subfloat[Sugarcane\label{fig:2e}]{\includegraphics[width=0.2\textwidth]{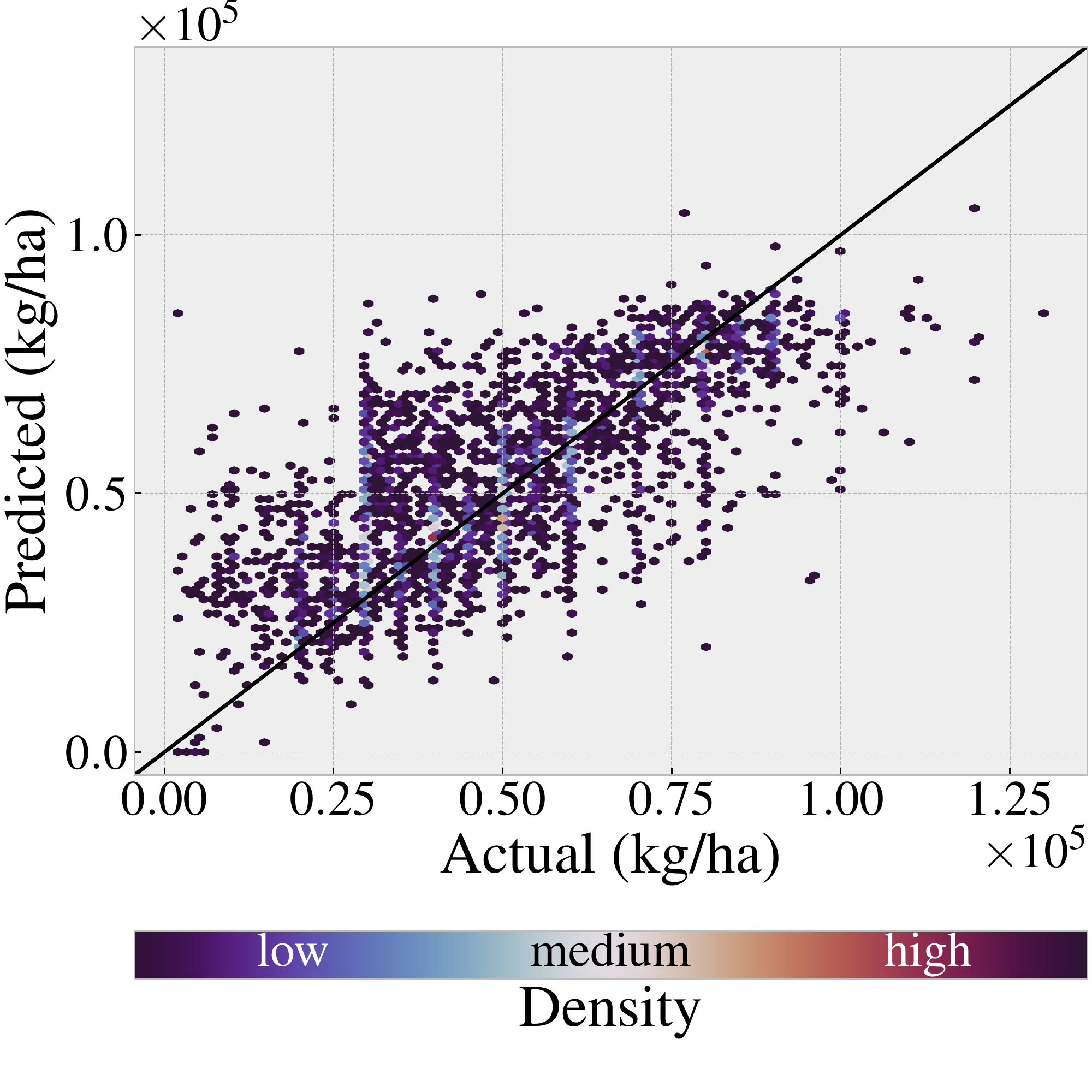}}\hfill
    \caption{
        Scatter plots showing the performance of the best trained models. The
        vertical axis shows the predicted values, while the horizontal axis
        shows the actual values. All evaluation points are related to crop
        production in Brazil in 2018. Predictions were only performed for
        municipalities that actually produced the crop in the evaluation year.
        Points were colored differently to show densities of regions in the
        plots.
    }\label{fig:2}
\end{figure*}

\begin{table}[tb]
    \centering
    \begin{tabular}{lrr}
\toprule
      Crop &  Train/Validation &  Test \\
\midrule
      Corn &             35079 &  5054 \\
    Cotton &              1816 &   223 \\
      Rice &             16378 &  1914 \\
   Soybean &             14190 &  2315 \\
 Sugarcane &             23860 &  3339 \\
\bottomrule
\end{tabular}
\caption{
        Dataset sizes for each crop. Sizes were obtained after removing
        municipalities with zero yield. The sets were split by adding
        yield data related to 2018 to the Test set and the remaining data
        to the Train/Valitation set.
    }
    \label{tab:data}
\end{table}

To evaluate the performance of our model, we decided to use the
\ac{pam}---Municipal Agriculture Production---dataset, a countrywide
Brazilian agriculture productivity dataset, made available the
\ac{ibge}\footnote{Available at
\url{https://sidra.ibge.gov.br/pesquisa/pam/tabelas}}. The \ac{pam} dataset
concentrates data about crop production in each municipality in Brazil and
makes available statistics such as area dedicated for planting, productivity
in metric tonnes per hectare, average productivity, among others. Although it
presents local data productivity, this information is aggregated for the
municipality. Therefore, it is not possible to know exactly when and where a
given crop was planted by using this dataset.

We evaluated our model using five different crops: corn, cotton, rice, soybean,
and sugarcane. To do so, we downloaded the average yield data in kg/ha from
the \ac{pam} dataset for all municipalities in Brazil for the years
2011-2018. Since Brazil has 5435 municipalities, the dataset ended up with
$5435 \times 8 = 43480$ entries per crop. Before feeding the data into the
model, we removed entries with zero productivity. Additionally, we separated
the data related to the year 2018 as a test set and used it only to evaluate
model performance (Table~\ref{tab:data}).

Using only the yield data is not enough to build a useful model. Hence, we
augmented the \ac{pam} data with weather data from the ERA5
reanalysis~\citep[][2016]{hersbach2016era5} dataset and soil type information from the
SoilGrids~\citep{hengl2017soilgrids250m} dataset, which has soil information (both observed and model-generated data) in a 250m grid for the whole planet. The weather data in
question are the maximum temperature, minimum temperature, accumulated
precipitation, and \ac{gdd} on a monthly basis. To decide which months to
gather data about, we used crop calendars provided by \ac{conab} for each
state-crop pair in Brazil. For example, in São Paulo state (SP), the
\ac{conab} document states that, for cotton, planting happens in the Spring
(from October to December) while harvest happens in the Fall (April to June).
Hence, we gather weather data from October to June ($n$ = 9 months) to build the
dynamic input to the model. Since a municipality is defined by a polygon but
we're building a point dataset, we assume that all points in a municipality
share the same features of its centroid, and we fetch data about it.

The time period in months we used for each crop are summarized in
Table~\ref{tab:crop-calendar}. Notice that even though we know the duration of
the plant-harvest cycle, it might start at different times (and even have
different lenghts) for different regions in a country and between countries.
In this paper, our area of study was limited to Brazilian cities, and we used
data published by \ac{conab}~\citep{conab2019acompanhamento}, summarized in
Table~\ref{tab:crop-calendar-states}. When data about a specific state was
missing, we used data about the region the state belongs to.

\begin{table}[tb]
    \centering
    \begin{tabular}{lr}
\toprule
      Crop & Time period (months) \\
\midrule
      Corn & 9  \\
    Cotton & 9  \\
      Rice & 8  \\
   Soybean & 9  \\
 Sugarcane & 12 \\
\bottomrule
\end{tabular}
\caption{
    Time range of the various crops studied in this paper based on crop
    calendar.
}\label{tab:crop-calendar}
\end{table}

With regards to the soil data, we used the seven layers of the SoilGrids
dataset for nine features: clay content, silt content, sand content, bulk
density, coarse fragments, cation exchange capacity, organic carbon content,
pH in $H_2O$ and pH in $K\mathrm{Cl}$. When combined with the latitude and
longitude of the centroid of the municipality, this yields the static input
of the model depicted in Figure~\ref{fig:architecture}.

To correctly assess the performance of the model, we performed thirty
independent executions of the train-test cycle. This is necessary due to the
intrinsic stochasticity of the training process, which uses random
initialization of the weight matrices of the neural network, as well as the
random shuffling of the mini-batches in the \ac{sgd} optimization process.
For training, we used the Adam~\citep[][2014]{kingma2014adam} algorithm with learning
rate $\alpha=5\times10^{-4}$, $\beta_1=0.9$, $\beta_2=0.999$, and a mini-batch
size of 280. The maximum number of training epochs was set to 500 with early
stopping set with a patience of 50. To reduce overfitting, we employed $L_2$ regularization, with $\lambda=1\times{}10^{-5}$.

To evaluate performance, we use two metrics: correlation and \ac{mape}.
Correlation gives a measure of how linear the relationship between predictions and
true values are, while \ac{mape} measures the relative error between true
values and predictions. We evaluated performance in 2018 because this is
closer to how the model would be used in practice: it would be trained with
all data available to make predictions for the current year. In our setting,
2018 represents this year, since it is the most recent one available in the
\ac{pam} dataset.

\begin{table}[tb]
    \centering
    \begin{tabular}{lrrrr}
\toprule
      Crop &  $\mu_{\text{Cor}}$ &  $\sigma_{\text{Cor}}$ &  $\mu_{\text{MAPE}}$ &  $\sigma_{\text{MAPE}}$ \\
\midrule
      Corn &               0.881 &                  0.003 &               47.855 &                   4.197 \\
    Cotton &               0.924 &                  0.003 &               29.238 &                   1.516 \\
      Rice &               0.874 &                  0.004 &               31.484 &                   1.194 \\
   Soybean &               0.300 &                  0.047 &               16.015 &                   0.969 \\
 Sugarcane &               0.710 &                  0.012 &               40.877 &                   1.656 \\
\bottomrule
\end{tabular}
\caption{
    Model performance with noise layer in dynamic features.
    $\mu_{\text{Cor}}$ represents the average value of correlation metric,
    $\sigma_{\text{Cor}}$ represents the standard deviation of the correlation metric. Similarly,
    $\mu_{\text{MAPE}}$ and $\sigma_{\text{MAPE}}$ represent the average and standard deviation of
    the \ac{mape}. Averages were computed over thirty independent executions.
}\label{tab:performance-noise}
\end{table}

\begin{table}[tb]
    \centering
    \begin{tabular}{lrrrr}
\toprule
      Crop &  $\mu_{\text{Cor}}$ &  $\sigma_{\text{Cor}}$ &  $\mu_{\text{MAPE}}$ &  $\sigma_{\text{MAPE}}$ \\
\midrule
      Corn &               0.865 &                  0.005 &               51.716 &                   6.114 \\
    Cotton &               0.920 &                  0.004 &               30.502 &                   1.586 \\
      Rice &               0.864 &                  0.005 &               32.826 &                   1.220 \\
   Soybean &               0.288 &                  0.025 &               14.755 &                   0.565 \\
 Sugarcane &               0.708 &                  0.012 &               42.081 &                   2.092 \\
\bottomrule
\end{tabular}
\caption{
    Model performance without noise layer.
    $\mu_{\text{Cor}}$ represents the average value of correlation metric,
    $\sigma_{\text{Cor}}$ represents the standard deviation of the correlation metric. Similarly,
    $\mu_{\text{MAPE}}$ and $\sigma_{\text{MAPE}}$ represent the average and standard deviation of
    the \ac{mape}. Averages were computed over thirty independent executions.
}\label{tab:performance}
\end{table}

The standard deviation of the noise layer was set to 0.3, and all the input
features were normalized to the 0--1 range prior to being input to the neural
network. Tables~\ref{tab:performance-noise} and~\ref{tab:performance} show the
performance of the model with and without this noise layer, respectively.  From
the tables, it can be seen that adding a noise layer improves model
performance, as correlation of the predicted yields for the model with noise
layer is higher for all crops, and \ac{mape} is smaller in almost all crops.

In Table~\ref{tab:performance-best}, we show the performance of the best
models found in the set of thirty for each crop. It is interesting to notice
that although cotton has the smallest dataset size (Table~\ref{tab:data}), it
is the one with best correlation. When we contrast this to Figure~\ref{fig:2},
which shows a scatter plot of the actual and predicted yields for 2018, we
conclude correlation is a good metric, since a visual inspection indicates
fit is good. For example, Figure~\ref{fig:2b} shows that there seems, indeed,
to be a good fit for cotton.

We see from Tables~\ref{tab:performance-noise} and~\ref{tab:performance} that
\ac{mape} for all crops is somewhat high, but upon close examination of
Figure~\ref{fig:2}, one is able to see various vertical lines in all the
scatter plots of crops. The vertical lines show that many municipalities have
\emph{exactly} the same productivity. We attribute this to the
nature of the \ac{pam} dataset constructed by \ac{ibge}, which relies on
self-reporting to compile the tables for crop production in Brazil. We
believe that an increase in the precision of the data
would greatly benefit the model itself and its predictions.

We also see that see that the best model for soybean
(Table~\ref{tab:performance-best}) has quite a low correlation, but low
\ac{mape}. Observing Figure~\ref{fig:2d} we see that soybean production seems
to have low bias, but high variance. Therefore, even though correlation is
low, the model makes predictions that are concentrated in the 2000--4000
kg/ha range, yielding the low \ac{mape} we observe.
For sugarcane, from the tables and from observing Figure~\ref{fig:2e} we see
the model has good predictive power, but is probably harmed because,
from the crop calendar we used, sugarcane can be planted on
any month and can be harvested in any month. Due to that, the features we
selected may not be powerful enough to explain the productivity, resulting
is harmed performance. Therefore, even though sugarcane has the second largest
dataset in this work (Table~\ref{tab:data}), the data itself is not enough to
yield a good predictor.

Rice and corn have similar correlations, but different \ac{mape} values. We
attribute this to the model underestimating the production of large corn
yields as shown in Figure~\ref{fig:2a}, which harms \ac{mape} for corn.
Dispersion in rice seems to be bigger in lower production values
(Figure~\ref{fig:2c}), resulting in a better \ac{mape}.

\begin{table}[tb]
    \centering
    \begin{tabular}{lrr}
\toprule
Crop      & $\max_{\text{Cor}}$ & $\min_{\text{MAPE}}$ \\
\midrule
Corn      & 0.886 & 41.596 \\
Cotton    & 0.929 & 26.648 \\
Rice      & 0.881 & 29.624 \\
Soybean   & 0.348 & 13.704 \\
Sugarcane & 0.739 & 37.304 \\
\bottomrule
\end{tabular}
\caption{
    Best performance of the model for each crop and metric.
    $\max_{\text{Cor}}$ represents the maximum correlation found over all
    executions for a given crop, while $\min_{\text{MAPE}}$ represents the
    minimum \ac{mape} found over all executions for a given crop. Maximum and
    minimum were computed over a set of thirty independent executions.
}\label{tab:performance-best}
\end{table}

\begin{table*}[tb]
  \centering
\begin{tabular}{llcc}
\toprule
                      &           & Planting        & Harvest       \\
\cmidrule{3-4}
State                 & Crop      &                 &               \\
\midrule
AC                    & Corn      & [01/10, 31/12]  & [01/02, 31/05]\\
                      & Cotton    & [01/12, 28/02]  & [01/05, 30/06]\\
                      & Rice      & [01/10, 31/12]  & [01/02, 30/04]\\
                      & Soybean   & [01/01, 15/06]  & [01/06, 31/10]\\
                      & Sugarcane & [01/01, 31/03]  & [01/04, 31/01]\\
AL                    & Corn      & [01/10, 31/01]  & [01/03, 30/06]\\
                      & Cotton    & [01/01, 28/02]  & [01/05, 30/11]\\
                      & Rice      & [01/10, 31/12]  & [01/01, 30/04]\\
                      & Soybean   & [01/10, 28/02]  & [01/03, 30/07]\\
                      & Sugarcane & [01/10, 31/08]  & [01/09, 30/03]\\
AM                    & Corn      & [01/04, 30/11]  & [01/08, 30/04]\\
                      & Cotton    & [01/12, 28/02]  & [01/05, 30/06]\\
                      & Rice      & [01/09, 31/12]  & [01/01, 31/12]\\
                      & Soybean   & [01/01, 15/06]  & [01/06, 31/10]\\
                      & Sugarcane & [01/01, 31/03]  & [01/04, 31/01]\\
AP                    & Corn      & [01/12, 31/01]  & [01/04, 31/05]\\
                      & Cotton    & [01/12, 28/02]  & [01/05, 30/06]\\
                      & Rice      & [01/10, 31/12]  & [01/02, 30/04]\\
                      & Soybean   & [01/01, 15/06]  & [01/06, 31/10]\\
                      & Sugarcane & [01/01, 31/03]  & [01/04, 31/01]\\
BA                    & Corn      & [01/10, 28/02]  & [01/03, 31/08]\\
                      & Cotton    & [01/11, 28/02]  & [01/05, 30/09]\\
                      & Rice      & [01/10, 31/01]  & [01/02, 30/06]\\
                      & Soybean   & [01/10, 31/01]  & [01/01, 30/04]\\
                      & Sugarcane & [01/01, 31/03]  & [01/04, 31/01]\\
CE                    & Corn      & [01/01, 30/04]  & [01/05, 31/08]\\
                      & Cotton    & [01/02, 30/04]  & [01/06, 31/08]\\
                      & Rice      & [01/02, 30/04]  & [01/05, 31/08]\\
                      & Soybean   & [01/10, 28/02]  & [01/03, 30/07]\\
                      & Sugarcane & [01/10, 31/08]  & [01/09, 31/03]\\
ES                    & Corn      & [01/10, 31/12]  & [01/02, 31/05]\\
                      & Cotton    & [01/11, 31/01]  & [01/04, 31/07]\\
                      & Rice      & [01/10, 31/12]  & [01/03, 30/06]\\
                      & Soybean   & [01/10, 31/12]  & [01/02, 31/05]\\
                      & Sugarcane & [01/01, 31/03]  & [01/04, 31/01]\\
GO                    & Corn      & [01/10, 31/12]  & [01/03, 30/06]\\
                      & Cotton    & [01/12, 31/01]  & [01/06, 31/08]\\
                      & Rice      & [01/10, 31/12]  & [01/03, 31/05]\\
                      & Soybean   & [01/10, 31/12]  & [01/01, 30/04]\\
                      & Sugarcane & [01/01, 31/03]  & [01/04, 31/01]\\
MA                    & Corn      & [01/11, 28/02]  & [01/03, 31/08]\\
                      & Cotton    & [01/12, 31/01]  & [01/07, 30/09]\\
                      & Rice      & [01/12, 31/03]  & [01/03, 30/06]\\
                      & Soybean   & [01/10, 28/02]  & [01/02, 31/07]\\
                      & Sugarcane & [01/10, 31/08]  & [01/09, 31/03]\\
MG                    & Corn      & [01/08, 31/12]  & [01/02, 30/06]\\
                      & Cotton    & [01/11, 30/04]  & [01/04, 30/09]\\
                      & Rice      & [01/10, 31/12]  & [01/03, 30/06]\\
                      & Soybean   & [01/10, 31/12]  & [01/01, 31/05]\\
                      & Sugarcane & [01/01, 31/03]  & [01/04, 31/01]\\
MS                    & Corn      & [01/09, 31/12]  & [01/02, 31/05]\\
                      & Cotton    & [01/10, 31/01]  & [01/04, 31/08]\\
                      & Rice      & [01/08, 31/12]  & [01/01, 30/04]\\
                      & Soybean   & [01/09, 31/12]  & [01/01, 30/04]\\
                      & Sugarcane & [01/01, 31/03]  & [01/04, 31/01]\\
MT                    & Corn      & [01/10, 31/12]  & [01/02, 30/06]\\
                      & Cotton    & [01/11, 31/01]  & [01/06, 30/09]\\
                      & Rice      & [01/10, 31/01]  & [01/01, 31/05]\\
                      & Soybean   & [01/09, 31/12]  & [01/01, 30/04]\\
                      & Sugarcane & [01/01, 31/03]  & [01/04, 31/01]\\
PA                    & Corn      & [01/10, 31/12]  & [01/02, 30/06]\\
                      & Cotton    & [01/12, 28/02]  & [01/05, 30/06]\\
                      & Rice      & [01/10, 31/12]  & [01/02, 30/04]\\
                      & Soybean   & [01/01, 30/04]  & [01/05, 31/08]\\
                      & Sugarcane & [01/01, 31/03]  & [01/04, 31/01]\\
\bottomrule
\end{tabular}
\quad
\begin{tabular}{llcc}
\toprule
                      &           & Planting        & Harvest       \\
\cmidrule{3-4}
State                 & Crop      &                 &               \\
\midrule
PB                    & Corn      & [01/01, 30/06]  & [01/07, 31/10]\\
                      & Cotton    & [01/02, 31/05]  & [01/06, 31/10]\\
                      & Rice      & [01/01, 31/03]  & [01/05, 31/07]\\
                      & Soybean   & [01/10, 28/02]  & [01/03, 30/07]\\
                      & Sugarcane & [01/10, 31/08]  & [01/09, 31/03]\\
PE                    & Corn      & [01/01, 31/05]  & [01/05, 31/08]\\
                      & Cotton    & [01/01, 28/02]  & [01/05, 30/11]\\
                      & Rice      & [01/01, 31/03]  & [01/05, 30/11]\\
                      & Soybean   & [01/10, 28/02]  & [01/03, 30/07]\\
                      & Sugarcane & [01/10, 31/08]  & [01/09, 31/03]\\
PI                    & Corn      & [01/12, 28/02]  & [01/04, 31/08]\\
                      & Cotton    & [01/12, 31/01]  & [01/06, 31/08]\\
                      & Rice      & [01/11, 28/02]  & [01/03, 30/06]\\
                      & Soybean   & [01/10, 31/01]  & [01/02, 30/06]\\
                      & Sugarcane & [01/10, 31/08]  & [01/09, 31/03]\\
PR                    & Corn      & [01/09, 31/12]  & [01/01, 31/05]\\
                      & Cotton    & [01/01, 28/02]  & [01/03, 30/06]\\
                      & Rice      & [01/08, 31/12]  & [01/01, 31/05]\\
                      & Soybean   & [01/09, 31/12]  & [01/01, 30/04]\\
                      & Sugarcane & [01/01, 31/03]  & [01/04, 31/01]\\
RJ                    & Corn      & [01/10, 31/12]  & [01/02, 31/05]\\
                      & Cotton    & [01/11, 31/01]  & [01/04, 31/07]\\
                      & Rice      & [01/10, 31/12]  & [01/03, 30/06]\\
                      & Soybean   & [01/10, 31/12]  & [01/02, 31/05]\\
                      & Sugarcane & [01/01, 31/03]  & [01/04, 31/01]\\
RN                    & Corn      & [01/02, 31/05]  & [01/06, 30/09]\\
                      & Cotton    & [01/02, 30/04]  & [01/06, 30/09]\\
                      & Rice      & [01/03, 31/05]  & [01/06, 30/11]\\
                      & Soybean   & [01/10, 28/02]  & [01/03, 30/07]\\
                      & Sugarcane & [01/10, 31/08]  & [01/09, 31/03]\\
RO                    & Corn      & [01/09, 31/12]  & [01/02, 31/05]\\
                      & Cotton    & [01/12, 28/02]  & [01/05, 30/06]\\
                      & Rice      & [01/10, 31/01]  & [01/02, 30/04]\\
                      & Soybean   & [01/10, 31/12]  & [01/01, 30/04]\\
                      & Sugarcane & [01/01, 31/03]  & [01/04, 31/01]\\
RR                    & Corn      & [01/04, 31/05]  & [01/09, 31/10]\\
                      & Cotton    & [01/07, 30/09]  & [01/02, 30/04]\\
                      & Rice      & [01/04, 30/06]  & [01/08, 31/10]\\
                      & Soybean   & [01/04, 30/06]  & [01/08, 31/10]\\
                      & Sugarcane & [01/01, 31/03]  & [01/04, 31/01]\\
RS                    & Corn      & [01/07, 28/02]  & [01/01, 30/06]\\
                      & Cotton    & [01/01, 28/02]  & [01/03, 30/06]\\
                      & Rice      & [01/09, 31/12]  & [01/02, 31/05]\\
                      & Soybean   & [01/10, 31/01]  & [01/02, 31/05]\\
                      & Sugarcane & [01/01, 31/03]  & [01/04, 31/01]\\
SC                    & Corn      & [01/08, 31/01]  & [01/01, 30/06]\\
                      & Cotton    & [01/01, 28/02]  & [01/03, 30/06]\\
                      & Rice      & [01/08, 30/11]  & [01/01, 31/05]\\
                      & Soybean   & [01/10, 30/12]  & [01/02, 31/05]\\
                      & Sugarcane & [01/01, 31/03]  & [01/04, 31/01]\\
SE                    & Corn      & [01/10, 31/01]  & [01/03, 30/06]\\
                      & Cotton    & [01/01, 28/02]  & [01/05, 30/11]\\
                      & Rice      & [01/08, 31/10]  & [01/12, 28/02]\\
                      & Soybean   & [01/10, 28/02]  & [01/03, 30/07]\\
                      & Sugarcane & [01/10, 31/08]  & [01/09, 31/03]\\
SP                    & Corn      & [01/10, 31/12]  & [01/03, 31/05]\\
                      & Cotton    & [01/09, 31/12]  & [01/04, 31/07]\\
                      & Rice      & [01/09, 31/12]  & [01/03, 31/05]\\
                      & Soybean   & [01/09, 31/12]  & [01/02, 31/05]\\
                      & Sugarcane & [01/01, 31/03]  & [01/04, 31/01]\\
TO                    & Corn      & [01/11, 31/01]  & [01/03, 30/06]\\
                      & Cotton    & [01/12, 28/02]  & [01/05, 31/07]\\
                      & Rice      & [01/11, 31/01]  & [01/01, 31/05]\\
                      & Soybean   & [01/10, 31/12]  & [01/02, 31/05]\\
                      & Sugarcane & [01/01, 31/03]  & [01/04, 31/01]\\
\bottomrule
\end{tabular}
   \caption{
    Planting and harvesting dates for each state in Brazil according to
    the crop calendars used in this study.
}\label{tab:crop-calendar-states}
\end{table*}
\section{Conclusion}\label{sec:conclusion}

This paper extended a deep-learning model from the
literature~\citep{oliveira2018scalable} to support five different crops in a
scalable manner. Different from previous work, which leverage satellite data
for direct farm observation for yield prediction, we employ weather and soil
data, which is computationally cheaper to process than large satellite
images. Additionally, by using our approach one can adopt weather
forecasts to predict yield before planting occurs, while using
\ac{ndvi}-based forecasts is only possible after plants reach a certain
growth stage.

Our results show that agriculture stakeholders can get insights into
potential yield even before planting to help the decision process when
dealing with farm and government activities like machinery rental,
contracting, price negotiation, and logistics planning. We provide accurate
results even with fewer data requirements for different crops in the presence
of noisy yield local data. Based on our results, it is possible to state that
our model is scalable enough to provide accurate predictions to the whole
world for different crops if a reliable local yield data is available. 
\normalsize
\bibliography{references} %

\end{document}